\documentstyle[aps,epsf,twocolumn,prl]{revtex}

\begin{document}
\draft

\twocolumn[\hsize\textwidth\columnwidth\hsize\csname @twocolumnfalse\endcsname
\title{The anomalous metallic ferromagnetic state of Sr doped manganites.
 }
\author{ V. Ferrari$^1$, M.J. Rozenberg$^1$ and R. Weht$^2$}

\address{$^1$Depto. de F\'{\i}sica, FCEN, Universidad de Buenos Aires,
Ciudad Universitaria Pab.I, (1428) Buenos Aires, Argentina.\\
$^2$ Depto. de F\'{\i}sica, CNEA, Avda. General Paz y Constituyentes,
(1650) San Martin, Argentina.}

\date{\today}
\maketitle
\widetext
\begin{abstract}
We deduce a model relevant for the anomalous metallic state of Sr doped
manganites at low temperatures within the ferromagnetic phase.
It provides a natural explanation to several anomalous features 
observed experimentally, such as the vanishing Drude contribution
in optical conductivity, the pseudo-gap in the density of states, an the
unusual dispersion observed in photoemission.

\end{abstract}
\noindent


]

\narrowtext

The fascinating physics of the perovskite manganite systems is
continuously attracting the interest of experimentalists and 
theorists alike. Initially, most of the attraction 
focused on the colossal magnetoresistance (CMR)
effect, due to its potential technological applications.
Nevertheless, these systems brought us other remarkable and perhaps more
puzzling surprises as, for instance,
the strange nature
of the metallic state at low temperatures.
Optical conductivity experiments in Sr doped manganites
La$_x$Sr$_{1-x}$MnO$_3$ by Tokura et al. \cite{tokura},
have demonstrated the presence of an extremely small Drude
part with a strong mid-infrared contribution
dominating the low frequency response. The dc-resistivity,
on the other hand, shows conventional $T^2$ metallic behaviour, but its
coefficient $A$ has a strong doping dependence, a feature 
typical of strongly correlated
systems \cite{urushibara}. In contrast, however, the specific heat experiments 
revealed a rather
small effective mass enhancement \cite{specheat}.
Recent X-ray photoemission studies on n-layer Sr doped
systems (n=2 and $\infty$) by Dessau et al. \cite{dessau}
showed additional puzzling aspects. For instance,
in the n=$\infty$ compound at $x=0.4$ there is a strong suppression
of spectral weight near the Fermi energy,
which correlates with the appearance of
a {\em pseudogap} in the n=2 bi-layer system.
Remarkably, this bi-layer system still shows a strong CMR effect
with a rather good metallic ferromagnetic (FM) state at low $T$.

Providing theoretical insight to understand
this puzzling combination is a challenging problem 
of strong current interest and the goal of the present
work. One possible approach would be to
assume that the anomalous behaviour is due to the combined interaction
of the many degrees of freedom which are {\it a priori} present in the
perovskite manganites, such as, spin, orbital and local lattice distortion
\cite{kaliulin,oles}.
Here we take a different stance and by
some reasonable simplifying assumptions, we
are lead to a minimal model Hamiltonian
that can be reliably solved. It will provide valuable
physical insight into the nature of the FM metallic state.

A rather general Hamiltonian for the low energy physics 
of the CMR manganite systems
is,\cite{millis,review}
\begin{eqnarray}
H  =   \sum_{\rm <ij>ab\sigma} t_{\rm ij}^{\rm ab} (c_{\rm ia\sigma}^\dagger 
c_{\rm jb\sigma}
    + h.c.) - J_H \sum_{\rm i} \vec{S}_{\rm i} \cdot \vec{s}_{\rm i} +
\nonumber \\
  H_{JT}  +  {U \over 2} \sum_{\rm iab\sigma\sigma'}
n_{\rm ia \sigma} n_{\rm ib \sigma'} (1 - \delta_{\rm ab} 
\delta_{\sigma \sigma'})
\label{hamil}
\end{eqnarray}
where $\rm a,b = 1,2$ are the orbital indexes of the $e_g$ bands,
$\sigma$ is the spin,
$t_{\rm ij}^{\rm ab}$ correspond to the amplitudes for nearest neighbours
hopping, and $\vec{S}$ denotes a local spin 3/2 due to the three electron 
in the $t_{2g}$ orbitals
The local spin density $\vec{s}$ of the ($1-x$) conduction 
electrons in the $e_g$ bands 
is coupled to the latter through the Hund's rule.
The first two terms define a ferromagnetic Kondo lattice
model and give a realization of the double exchange mechanism (DEM)
\cite{zener,kubo}, 
the third includes Jahn-Teller polaronic effects \cite{millis2}. 
These terms are widely considered to be responsible for much of
the physics associated to the CMR effect that occurs at intermediate
$T$ of the order of the ferromagnetic transition $T_c$.
However, it was recently emphasized that the model also predicts
phase separation \cite{moreo}
in various regions of parameter space, which may 
play an important role for the complete understanding of the CMR.
The last term accounts for local
correlations due to a Hubbard-like interaction. 
While this interaction is recognized as
an essential ingredient in every other transition metal oxide,
it has only
received very limited attention in manganites so 
far \cite{ishihara,mjr,imada}.
As we shall see, it
will play a most important
role in the anomalous FM phase
at low $T$. 

The exact solution of the full model (\ref{hamil}) 
is clearly not possible, so we shall
make a few simplifying assumptions which are well justified
in the FM phase at low enough temperatures.
In passing we recall that the stability of the FM state is
due to the DEM, that favors the gain in kinetic energy
when the system is doped with $x$ holes per site \cite{degennes,furukawa}.
In a first approximation, we can neglect the
Jahn-Teller polaronic term $H_{JT}$ since the dynamical local distortions
of the lattice, which have a notable effect in insulating phases
and close to $T_c$, are found to disappear when one enters the
metallic FM state \cite{radaelli,martin,millis}. 
On the other hand, we can also safely 
assume the local spins 3/2 to be uniform and static and the
magnetization of the $(1-x)$
conduction electrons to be perfectly aligned to them.
This assumption is justified by the experimental observation
that at low enough temperatures (below $\sim$200K at $x \sim 0.2$)
the local magnetization is found to saturate at the classical value
\cite{urushibara}.
In consequence, in (\ref{hamil}) we can further drop
all terms containing operators labeled with, say, the
down spin index $\downarrow$.
At this point, we may take the limit of large lattice connectivity
(or $d \to \infty$) and use the 
dynamical mean field theory (DMFT) \cite{metzvoll,review}
as a controlled approximation to solve our model.
The lattice Hamiltonian is mapped onto an 
impurity problem with a self-consistency condition, and
the resulting local action reads,

\begin{eqnarray}
S_{\rm loc}= \sum_{\rm nab}  d_{\rm a}^\dagger ( i\omega_{\rm n}) 
[G_0^{-1}( i\omega_{\rm n})]_{\rm ab} d_{\rm b}(i\omega_{\rm n})
\nonumber\\
+ \beta {U \over 2} 
\sum_{\rm n a \ne b} n_{\rm a}(i\omega_{\rm n}) n_{\rm b}(-i\omega_{\rm n})
\label{action}
\end{eqnarray}
\noindent
where $[G_0^{-1}(i\omega_{\rm n})]_{\rm ab}$ obeys
the self-consistency condition
$
[G_0^{-1}(i\omega_{\rm n})]_{\rm ab} = (i\omega_{\rm n} + \mu)
\delta_{\rm ab} + \Sigma_{\rm a'b'} t^{\rm ab'} 
[G(i\omega_{\rm n})]_{\rm b'a'} t^{\rm a'b}
$
with $G$ the local Green's function
\cite{note2}
For realistic $e_g$-bands the off diagonal hopping elements
of the matrix $t_{\rm ij}^{\rm ab}$ are of the same order of magnitude as the 
diagonal ones, thus, for simplicity, we may set $t_{\rm ij}^{\rm ab} = -t$.
With a rotation of the impurity operators  
$\{d_{\rm 1},d_{\rm 2}\} \to \{\gamma=(d_{\rm 1}+d_{\rm 2})/ \sqrt{2}, 
\gamma'=(d_{\rm 1}-d_{\rm 2})/ \sqrt{2}\} $, the matrix
$G_0^{-1}(i\omega_{\rm n})$ becomes
diagonal and the self-consistency condition now reads,
\begin{equation}
G_0^{-1}(i\omega_{\rm n}) = 
\left[\begin{array}{cc} i\omega_{\rm n}+\mu - (2t)^2 G(i\omega_{\rm n})
 & 0 \\
0 & i\omega_{\rm n}+\mu \end{array}\right]
\label{self2}
\end{equation}

This equation tells us that in the new basis
the $\gamma-$electrons are mobile with
hopping amplitude $2t$, while the $\gamma'-$electron are localized.
As the interaction term in (\ref{action}) is rotationally invariant,
the self-consistent system of equations 
(\ref{action}) and (\ref{self2}) becomes formally
identical to that of a Falicov-Kimball model (FKM) \cite{falicov,brand}.
In this analogy, the  orbital degrees of freedom labeled by $\gamma$
and $\gamma'$ correspond to the light and heavy spin-less electrons
of the FKM.

\begin{figure}
\epsfxsize=3.5in
\epsffile{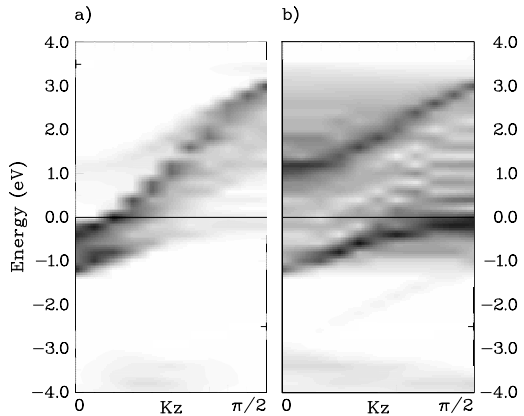}
\caption{
LDA bands for majority spin Mn orbitals of FM La$_{1-x}$A$_x$MnO$_3$
with x=0.175. The results are for bands with 
$d_{z^2}$ (a) and  $d_{x^2-y^2}$ (b) character along the $K_z$ direction.
The intensity corresponds to the average over 
$K_x$ and $K_y$ in the Brillouin zone.
Doping is similuted in the Virtual Crystal Approximation.
}
\label{fig1}
\end{figure}

Before discussing 
the physics of the model in the present context, we would
like to show that after the various manipulations and simplifying
assumptions made we have not lost touch with reality, and on the contrary,
we did gain precious insight.
To this end, we use density functional theory and obtain the realistic
bandstructure of ferromagnetic LaSrMnO$_3$ within the local density 
approximation. The bandstructure for the majority spin electrons is
plotted in Fig.\ref{fig1} 
along one of the main directions of the cubic structure.
The results shown correspond to the two $Mn$ bands with $e_g$ symmetry
averaged over the Brillouin zone. It can be clearly observed in the figure 
that the $d_{z^2}$ band is dispersive and crosses the Fermi energy, while
the $d_{x^2-y^2}$ band near the Fermi energy is essentially flat.
The results of this realistic calculation give strong support to the
simplified model that we derived before. 
We have therefore gained valuable physical insight as it will turn out
that the anomalous metallic behavior is due to the strong scattering
of fast electrons moving in the $d_{z^2}$ channel colliding with the 
very slow electrons in the $d_{x^2-y^2}$ channel.
Indeed, this is the basic mechanism that leads to non-Fermi liquid behavior
in the DMFT solution of the FKM \cite{skg,goetz}.
This effect occurs near half filling (in the so called ``pinning'' region)
and the incoherent metallic behavior can be thought of arising from
the superposition of the upper and lower Hubbard bands
that are split by the interaction strength $U$.

We shall now show that many anomalous properties of the FM metallic state
observed in $Sr-$doped manganites are qualitatively captured by
our model, using realistic values for the parameters.
For the particular case of the FKM it is well known \cite{skg,goetz,furukawa}
that the qualitative behavior of the spectral functions does not
depend on the choice of the non-interacting density of states (DOS).
Therefore, we shall use a Lorentzian DOS \cite{skg,furukawa}
that allow for fast and simple calculations. 
In our units $2t$=1eV, 
that gives a bandwidth W(FWMH)=2eV,
as suggested by bandstructure calculations.
The value of the local Coulomb repulsion is set to $U$=3eV, which is 
consistent with the experimental
estimates of either the charge transfer or the Mott gap
in Ref.\cite{park}.

\begin{figure}
\epsfxsize=3.5in
\epsffile{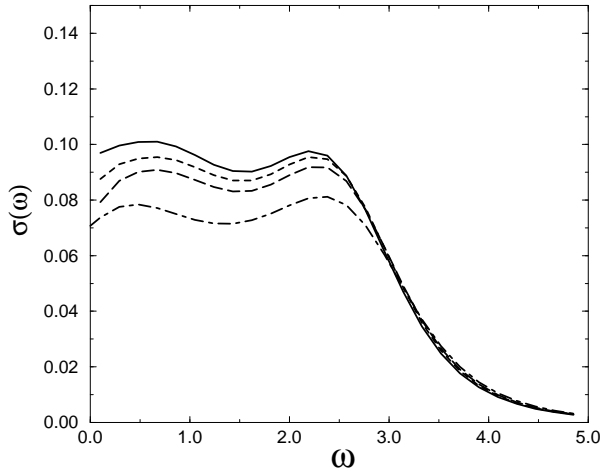}
\caption{
Optical conductivity $\sigma(\omega)$ calculated for various temperatures
$T/W$=0.005, 0.015, 0.05, 0.15 (solid, dashed, long dashed, dot-dashed)
and total number of particles $n$=0.93. 
}
\label{fig2}
\end{figure}

In Fig.\ref{fig2} we show the results for the optical 
conductivity \cite{goetz}
at various temperatures and small doping \cite{note1}. 
We clearly observe the lack
of a Drude part which is due to the non-Fermi liquid character of
the metallic state. This Drude contribution is absent even at $T=0$
and we can understand this result as due to the presence of localized
electrons that break the translational invariance. 
More physically, the incoherent metallic state is due to the
strong scattering of itinerant electrons at the
positions occupied by the localized ones.  We thus
find that the optical spectra is dominated by the mid-infrared contribution
which has a $T$ dependence very similar to the experimental results of
Okimoto et al. \cite{tokura}. Note that the lower
values of $T$ were chosen to be of the
same order of magnitude as in the experiments.

\begin{figure}
\epsfxsize=3.5in
\epsffile{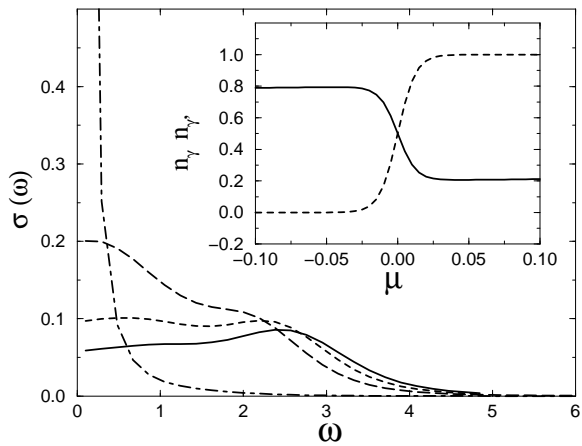}
\caption{
Optical conductivity $\sigma(\omega)$ for $T/W=0.005$ and
the total number of particles $n=$0.95 (solid),
0.93 (dashed), 0.9 (long dashed) and 0.8 (dash-dotted).
The inset shows the number 
occupation $n$ of the $\gamma$ (solid) and $\gamma '$ (dashed)
orbitals as a function of the chemical potential $\mu$.
}
\label{fig3}
\end{figure}
 
In Fig.\ref{fig3} we plot the results for the optical response
as we doped the system away from half filling. We observe that
the Drude part contribution rapidly emerges as the model recovers 
its translational invariance and the metallic state
looses its non-Fermi liquid character. 
As we show in the inset of the figure, 
the reason for this behaviour is that
when we reduce
the chemical potential $\mu$ (decreasing the total number
of particles) the occupation of the localized
$\gamma '$-electrons rapidly decreases 
and, at the same time,
the occupation of the itinerant $\gamma$-electrons {\em increases}.
Thus we have the simultaneous effects of a reduction of the scattering 
and an increase in the number of carriers with doping.
We would like to point out, that this behavior
is in good qualitative agreement with recent experimental
results of Tokura et al.\cite{tokura2}.

\begin{figure}
\epsfxsize=3.5in
\epsffile{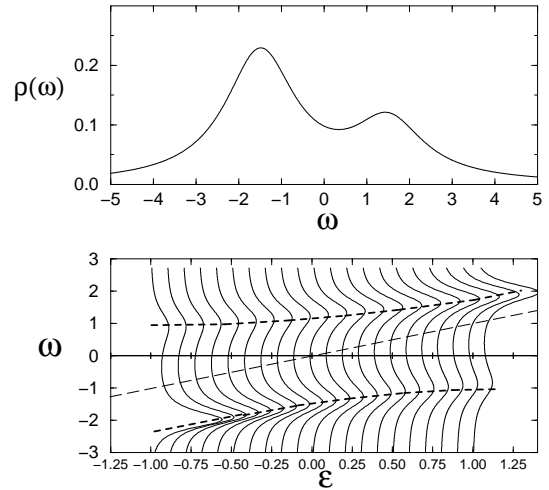}
\caption{
The top figure is the density of states $\rho(\omega)$ obtained for 
$T$=0.005 and $n$=0.93.
The bottom figure shows the
$\epsilon$-resolved density of states $\rho_\epsilon(\omega)$ calculated
for various values of $\epsilon$ across the Brillouin zone. Note
that the vertical axis correspond to the frequency $\omega$ and the
the curves are plotted along the horizontal axis to better display
the dispersion of the quasiparticle feature. The spectra
$\rho_\epsilon(\omega)$ were
shifted (in the horizontal direction) by their corresponding $\epsilon$ value. 
The thick dotted line is an aid to the eye to follow the 
dispersion of the quasiparticle feature as a function of $\epsilon$ across
the Brillouin zone. The thin dashed line shows the itinerant $\gamma$-electron
band dispersion of the non interacting FK model.
}
\label{fig4}
\end{figure}
 
We now turn to the calculation of the density of states (DOS) which allows
for the interpretation of the remarkable X-ray photoemission 
results \cite{dessau}.
The photoemission experiments were performed on 
$n-$layer systems at $x=0.4$.  The $n$=2 is a bi-layer system 
and the $n$=$\infty$ corresponds
to the usual 3-dimensional compound. In the first case the samples
can be cleaved and angle resolved photoemission results were obtained.
They revealed the unexpected feature of a well defined quasiparticle resonance
that shows usual band dispersion behavior away from the Fermi surface,
but the dispersion flattens out as it approached the (expected
location of the) Fermi surface, and remains flat without crossing
$\omega$=0. After angle integration, this behavior leads to
the presence of a so called {\em pseudogap} at the Fermi energy.
In the $n$=$\infty$ compound, angle resolved experiments are not
possible and the angle integrated photoemission also shows a strong
suppression of the spectral weight at $E_F$.
In the upper part of Fig.\ref{fig4} we show the 
model prediction for the DOS that clearly show
a suppression of spectral weight near $\omega$=0. This is due
to the splitting of the upper and lower Hubbard bands by an energy of
order $U$.
To gain further insight on the nature of the Hubbard bands and to
make comparison of the predictions of the model to the angle resolved
data, we compute the analog to a momentum resolved DOS within
DMFT. In this theory \cite{metzvoll,review} 
the Green's function implicitly depends on momenta 
through the single particle energy $\epsilon \equiv \epsilon_{\vec{k}}$ as,
$
G(\epsilon,\omega) = 1 /( \omega - \epsilon - \Sigma(\omega))
$
where $\Sigma(\omega)$ is the local self-energy.
Therefore, the connection to the usual $\vec{k}$-resolved DOS is made by 
noting that
momenta close to the zone center ($\Gamma$-point) correspond
to values of  $\epsilon \sim -W$; momenta  near
$E_F$, to values of $\epsilon \sim 0$; and momenta close
to the zone boundary, to
$\epsilon \sim W$.
In the lower part of 
Fig.\ref{fig4} we plot the results for the imaginary part of
$G(\epsilon,\omega)$ for various values of $\epsilon$ across the Brillouin 
zone. These results provide insights on the nature of the dispersion
of the interacting model and
can be qualitatively compared to the angle resolved
data of Ref.\onlinecite{dessau}.
Using the same realistic values for the model parameters, we clearly see in 
the figure a distinct quasiparticle
feature that disperses across the lower Hubbard band
in the occupied part of the spectra ($\omega < 0$), 
and another one that has similar
behaviour in the upper Hubbard band for the unoccupied part ($\omega > 0$).
The first one, which is to be contrasted to photoemission experiments,
shows a rather usual dispersive pattern
for $\epsilon \sim -W$ (near the zone center), but it
then flattens out as $\epsilon \to 0$, and remains flat up to higher values of
$\epsilon$ (near the zone boundary).
It is important to note that as the quasiparticle peak
fails to cross the Fermi 
energy, it also becomes broader and has a rapid loss of weight,
indicating its progressive incoherent
nature as it approaches  the low frequencies $\omega \sim 0$.
This behavior is in remarkable qualitative agreement with the angle resolved 
data in the bi-layer system \cite{dessau}. 

To conclude,
we have derived and studied a model relevant for the anomalous
FM metallic state of Sr doped manganites at low $T$.  Our results
provide a consistent description of several unusual observations in the
experimental spectroscopy. The analogy of our model
with an {\em orbital} Falicov-Kimball model, also explains the experimental
finding that at upon doping to high values of $x$, the 
incoherent metal crosses-over to 
a regular Fermi liquid. Even more interestingly,
at low values of doping, the analogy with the FKM
suggest also an explanation for the driving force in
the recent observation by Endoh {\it et al.}
\cite{endoh}
of charge-ordering in the FM {\em insulating} 
phase as due to the well known
charge density wave instability \cite{brand,vandong}
of the model in bipartite lattices near half filling. 
We stress that our present work remains at a 
qualitative level, and a detailed calculation of
the phase boundaries \cite{freericks}
and the eventual modifications that one may have to
introduce to make more precise comparison with experiments
are left for future work.

We acknowledge valuable discussions with A. Millis, B. Shraiman,
Y. Tokura and A.
Lichenstein. VF acknowledges support of FOSDIC.
MJR acknowledges support of Fundaci\'on Antorchas,
CONICET (PID $N^o4547/96$), and ANPCYT (PMT-PICT1855).


\begin{thebibliography}{99}
\bibitem{note1}
In DMFT, $\sigma(\omega)$ is formally derived for a Gaussian DOS and
only depends on the local $\Sigma(\omega)$. 
The $\Sigma(\omega)$ calculated here with a Lorentzian DOS
have the qualitatively correct behavior \cite{skg} which
directly implies the correct behavior of $\sigma(\omega)$.
For instance, the sum-rules are fullfilled within few percents.
The same procedure was also used in \cite{millis,furukawa}.
\bibitem{note2}
The self-concistency condition is derived within
the retraceable path approximation and is exact for a
Bethe lattice.
\end{thebibliography}
\end{document}